\documentclass[%
 twocolumn, amsmath, amssymb, aps, nofootinbib
]{revtex4-2}

\usepackage{graphicx}
\usepackage[dvipsnames]{xcolor}
\usepackage{dcolumn}
\usepackage{bm}
\usepackage{braket}

\usepackage[colorlinks=true,
    linkcolor=blue,
    filecolor=blue,
    citecolor=blue,      
    urlcolor=blue,]{hyperref}
\usepackage{mhchem}
\usepackage{multirow}

\usepackage{prettyref}
\newcommand{\pref}[1]{\prettyref{#1}}
\newrefformat{fig}{Fig.~\ref{#1}}
\newrefformat{tab}{Table~\ref{#1}}
\newrefformat{sec}{Sec.~\ref{#1}}
\newrefformat{eq}{Eq.~(\ref{#1})}

\begin{document}

\title{Multipoles as quantitative order parameters for altermagnetic spin splitting}

\author{Francesco Martinelli}\email{fmartinelli@ethz.ch}
\author{Anouk Droux}
\author{Claude Ederer}\email{edererc@ethz.ch}
\affiliation{Materials Theory, ETH Z\"{u}rich, Wolfgang-Pauli-Strasse 27, 8093 Z\"{u}rich, Switzerland}

\date{\today}

\begin{abstract}
We establish a quantitative relation between the altermagnetic spin-splitting and different higher order multipoles of the charge and magnetization density around the magnetic atoms. Magnetic multipoles such as octupoles or triakontadipoles have been suggested as potential ferroic order parameters for $d$- and $g$-wave altermagnetism, respectively, based mainly on qualitative symmetry arguments. We use first-principles-based electronic structure calculations to establish a clear quantitative relation between the strength of the altermagnetic spin splitting and the magnitude of certain local multipoles. We vary the magnitude of these multipoles either by applying an appropriate constraint on the charge density or by varying a corresponding structural distortion mode, using two simple perovskite materials, SrCrO$_3$ and LaVO$_3$, as model systems. Our analysis indicates that in general the altermagnetic spin splitting is not exclusively determined by the lowest order nonzero magnetic multipole, but results from a superposition of contributions from different multipoles with comparable strength, suggesting the need for a multi-component order parameter to describe altermagnetism. We also discuss different measures to quantify the overall spin-splitting of a material, without relying on features that might be specific to only individual bands.
\end{abstract}

\maketitle


\section{Introduction}

It has been recognized only very recently, that in certain antiferromagnets, now termed \emph{altermagnets}, time reversal symmetry is broken such that the Kramer's degeneracy of the electronic states is lifted even in the limit of zero spin-orbit coupling, while the net magnetization remains zero~\cite{Hayami/Yanagi/Kusunose:2019, Ahn_et_al:2019, Yuan_et_al:2020, Hayami/Yanagi/Kusunose:2020, Smejkal_et_al:2020, Yuan_et_al:2021, Smejkal/Sinova/Jungwirth:2022, Smejkal/Sinova/Jungwirth:2022a, Yuan/Zunger:2023, Yuan_et_al:2023, Guo_et_al:2023, Zeng/Zhao:2024, Lee_et_al:2024, Krempasky_et_al:2024, Reimers_et_al:2024, Aoyama/Ohgushi:2024, Lin_et_al:2024, McClarty/Rau:2024, Bai_et_al:2024, Jungwirth_et_al:2024, Cheong/Huang:2024, Radaelli/Gurung:2025, Hu_et_al:2025}.
This means that, for a general $\mathbf{k}$-point in the Brillouin zone of an altermagnet, the spin degeneracy of the electronic bands is lifted, but the presence of certain symmetry operations requires that this spin-splitting is reversed in other regions of the Brillouin zone, such that integrals over the whole Brillouin zone do not carry any net spin dependence.

Altermagnets therefore combine aspects of both ferromagnetic and antiferromagnetic (AFM) materials. Their antiferroic (AF) arrangement of magnetic dipoles is robust to stray magnetic fields and allows for faster dynamics, while the spin-splitting in their band-structure enables $\mathbf{k}$-dependent spin transport, with efficient spin-current generation recently demonstrated~\cite{Mazin/ThePRXEditors:2022, Naka_et_al:2019, Ma_et_al:2021, Gonzalez-Hernandez_et_al:2021, Shao_et_al:2021, Bose_et_al:2022}.
Moreover, in addition to potential applications in conventional spintronics, a plethora of other interesting properties of altermagnets have been suggested, such as a spontaneous Hall effect~\cite{Smejkal_et_al:2020, Smejkal_et_al:2022, Reichlova_et_al:2021, Feng_et_al:2022, GonzalezBetancourt_et_al:2023, Sato_et_al:2024, Cheong/Huang:2024}, multipole transport~\cite{Ko/Lee:2025, Han_et_al:2025, Baek/Han/Lee:2025}, unconventional superconducting properties~\cite{Mazin:2025, Chakraborty/Black-Schaffer:2024, Zhang/Hu/Neupert:2024, Zhu_et_al:2023, Banerjee/Scheurer:2024}, piezomagnetism~\cite{Bhowal/Spaldin:2024, Radaelli:2024}, chiral magnon splitting~\cite{Smejkal_et_al:2023, Liu_et_al:2024}, and many others
\cite{Mazin_et_al:2021, Verbeek_et_al:2024, Yershov_et_al:2024, Yang_et_al:2025, Smejkal:2024, Gu_et_al:2025, Nag_et_al:2024}.

On the theoretical side, much work has been focused on clarifying the symmetry aspects of altermagnets, in terms of the underlying spin-symmetry groups~\cite{Litvin/Opechowski:1974, Litvin:1977, Smejkal/Sinova/Jungwirth:2022a, Smejkal/Sinova/Jungwirth:2022, Urru_et_al:2025}, where, in the absence of spin-orbit coupling, rotations in spin-space are decoupled from rotations in real space. Thereby, the presence of a non-relativistic spin-splitting (NRSS) requires the absence of symmetry elements that map one spin sub-lattice on the other via simple translations or inversion, while the presence of a symmetry operation that maps the two sublattices on each other via a proper or improper rotation ensures a vanishing net magnetization~\cite{Smejkal/Sinova/Jungwirth:2022, Smejkal/Sinova/Jungwirth:2022a}.
However, a general quantitative description that goes beyond symmetry arguments, and relates the overall magnitude of the NRSS to a simple order parameter, remains to be fully established.

In a ferromagnet, the relevant order parameter is its net magnetic dipole moment, or magnetization, and the average spin-splitting is quantitatively related to this order parameter. For the case of altermagnets, higher order magnetic multipole moments, specifically magnetic octupoles or triakontadipoles, have been suggested as suitable ferroic order parameters~\cite{Bhowal/Spaldin:2024, McClarty/Rau:2024}, since the symmetries that allow these magnetic multipoles to be nonzero are identical to the symmetries that allow for a NRSS. 

Magnetic multipoles characterize the spatial distribution of the magnetization density around the magnetic atoms, and can potentially provide a simple and systematic unifying framework for classifying ferroically ordered magnetic states.
Thereby, a non-vanishing net magnetic dipole corresponds to ferromagnetic order, while altermagnetic states would correspond to a zero magnetic dipole but non-vanishing higher order multipoles. The lowest order non-vanishing magnetic multipole then determines the $\mathbf{k}$-space symmetry of the NRSS, with, e.g., $d$-wave and $g$-wave patterns corresponding to magnetic octupoles and triakontadipoles, respectively~\cite{Urru_et_al:2025}. 

Nevertheless, while the general relation between magnetic multipoles and altermagnetic symmetry has been pointed out several times~\cite{Bhowal/Spaldin:2024, Verbeek_et_al:2024, Nag_et_al:2024, McClarty/Rau:2024}, a clear quantitative correspondence between the magnitude of these multipole moments and the strength of the altermagnetic spin splitting still needs to be established.
Apart from its fundamental relevance, an appropriate definition of an altermagnetic order parameter allows for a quantitative comparison across different materials and can also enable a more targeted search for ``strong'' altermagnets with optimized properties. Furthermore, identification of an appropriate order parameter might also allow to control the phase transition and domain formation via its conjugate field. 

Here, we explore the quantitative relation between the overall strength of the altermagnetic NRSS and the magnitude of different charge and magnetic multipoles, using electronic structure calculations based on density functional theory (DFT). To this end, we vary the magnitude of specific multipoles by constraining the underlying electron density accordingly~\cite{Schaufelberger_et_al:2023}, or by systematically imposing different structural distortion modes. We then monitor the resulting changes in the NRSS as well as the corresponding $\mathbf{k}$-space symmetry.

We use two different materials as ``model systems'' to establish a quantitative relation between multipoles and NRSS. First, SrCrO$_3$, with a simple perovskite structure and C-type AFM order, which allows us to investigate the emergence of a NRSS driven by a purely electronic symmetry breaking, achieved by applying suitable perturbations to constrain specific multipoles~\cite{Schaufelberger_et_al:2023}. And second, LaVO$_3$, as a representative of a structurally distorted perovskite exhibiting the most commonly observed, so-called GdFeO$_3$-type distortion, which lowers the space group symmetry to $Pbnm$ and involves collective rotations of the oxygen octahedra surrounding the magnetic cations.
As recently pointed out~\cite{Bandyopadhyay/Picozzi/Bhowal:2025}, the corresponding distortions lead to different multipoles that allow for the emergence of a NRSS in combinations with either A-, C-, or G-type AFM order.

Our analysis shows that in many cases a clear quantitative relation, where the NRSS indeed scales with the magnitude of specific multipoles, can be established, but that in some cases the NRSS also emerges from a superposition of different components, with dominant contributions also from higher order multipoles than the lowest order nonzero one. 
Furthermore, we also assess strengths and limitations of different measures for the overall NRSS, to ensure a reliable quantitative analysis.

\section{Theoretical background and computational method}

In this section, we first summarize the definition of multipoles of the charge and magnetization density, and also provide the details of how we evaluate and constrain local multipoles. We then define the measures we use to quantify the NRSS, and finally we provide all other relevant computational details.

\subsection{Multipoles and how to constrain them within DFT}

\begin{figure}
    \includegraphics[width=0.35\textwidth]{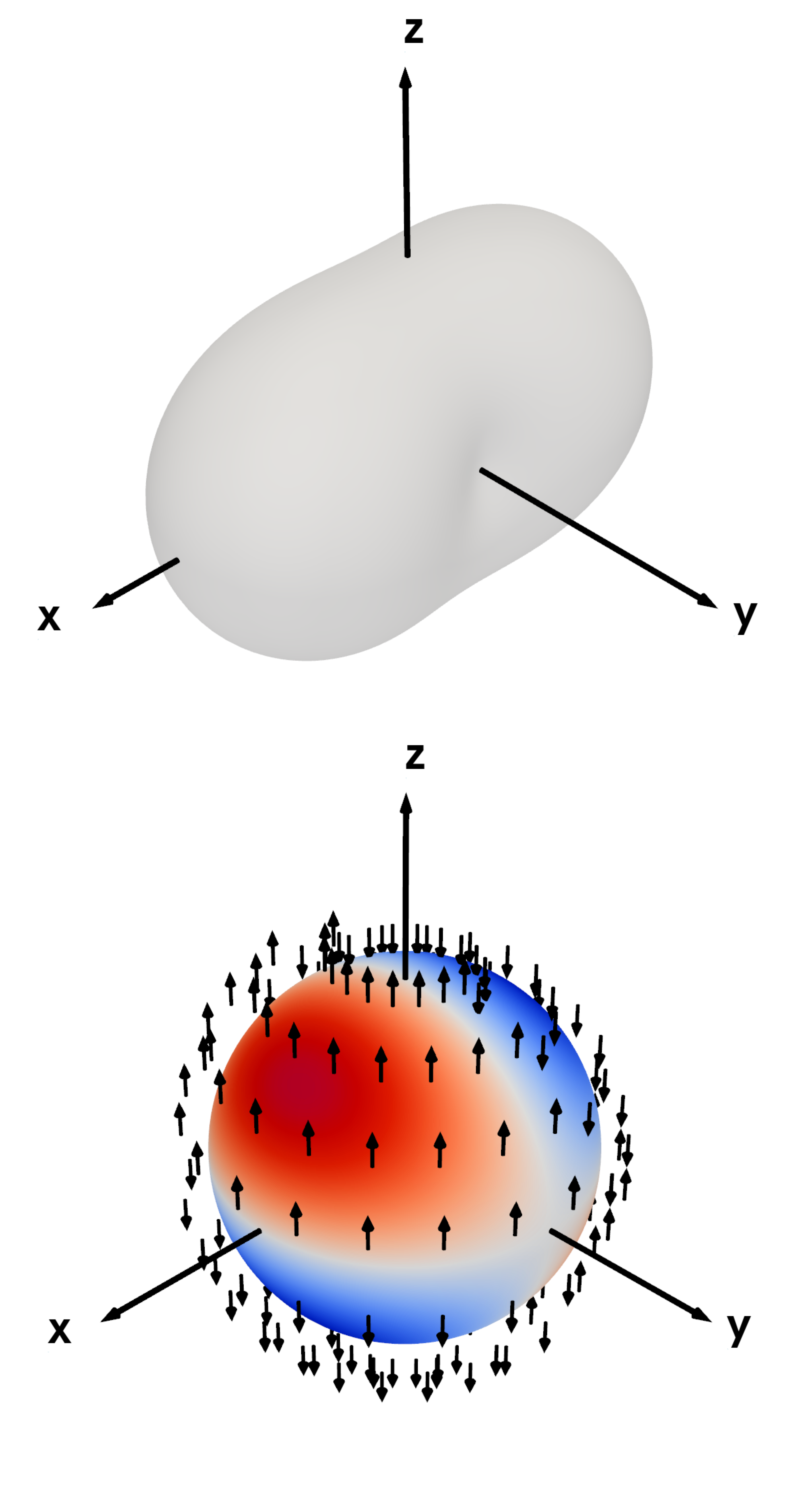}
    \caption{Top: Sketch of a charge density exhibiting a nonzero $\mathcal{Q}_{x^2 - y^2}$ quadrupole, characterized by charge accumulation (depletion) along the $x$ ($y$) axis. Bottom: Sketch of a collinear magnetization density with non-zero $\mathcal{O}_{xz}$ octupole. In the red (blue) regions the magnetization points upwards (downwards).}
    \label{fig:multipolescartoon_nl}
\end{figure}

Multipoles provide a systematic parametrization of the spatial dependence of the charge and magnetization densities around an atom (or other center). 
To treat charge and magnetic multipoles on the same footing, we can define a generalized density, 
$\varrho^p_q(\mathbf{r})$ (where $p \in \{0,1\}$ and $q=-p, \dots, p$) as follows: 
\begin{equation}\label{eq:0}
\mathrm{\varrho}^p_q (\mathbf{r}) = \mathrm{Tr}_{s} \left[ \langle \mathbf{r} | \hat{\sigma}^p_q \hat{\rho} | \mathbf{r} \rangle \right] \quad .
\end{equation}
Here, $\hat{\rho}$ is the density operator, $\hat{\sigma}^p_q$ represents either the identity operator in spin space (for $p=0$) or the operator corresponding to the Pauli matrices (for $p=1$), and the trace is taken over the spin degrees of freedom. 
Thus, $\varrho^0_0(\mathbf{r})$ is equal to the charge density, $\rho(\mathbf{r})$, while $\varrho^1_q$ defines the components of the magnetization density $\mathbf{m}(\mathbf{r})$. 

A general multipole of spatial order $k$ can then be defined as the integral over a certain component of the density multiplied with $k$ factors of the form $r_i$:
\begin{equation}
\label{eq:multipole-cartesian}
\mathcal{M}^{k, p}_{ij\dots l, q} = \int \underbrace{r_i r_j \cdots r_l}_{\text{$k$ such factors}}\, \varrho^p_q(\mathbf{r}) d\mathbf{r} \quad ,
\end{equation}
where $r_i$ is a cartesian component of the vector $\mathbf{r}$. Note that in principle the full spatial depdendence of the density can be reconstructed from the full set of multipole moments (with $k = 0, 1, \dots, \infty$). Such multipoles also appear naturally if one considers the interaction energy of a charge and magnetization density with an inhomogeneous electrostatic potential and a magnetic (Zeeman) field, where each multipole then interacts with a specific component or derivative of the corresponding field (see, e.g., Refs~\cite{Bhowal/Spaldin:2021, Schaufelberger_et_al:2023,  Bandyopadhyay/Picozzi/Bhowal:2025}).

Expressing the integrand in \pref{eq:multipole-cartesian} in spherical coordinates leads to a definition of spherical multipoles as:
\begin{equation}
\label{eq:spherical-multipole}
\mathcal{M}^{kp}_{hq} = \int r^k \, Y^k_h(\theta,\phi) \, \varrho^p_q(r,\theta,\phi) \, d\mathbf{r} \quad .
\end{equation}
These spherical multipoles can then be transformed to fully irreducible spherical tensors of rank $r$, with \mbox{$r=|k-p|, \dots, k+p$}, by coupling the spatial and spin indices~\cite{Sakurai_Napolitano:2020, Santini_et_al:2009, Bultmark_et_al:2009}:
\begin{equation}
w^{kpr}_t = \sum_{\substack{h=-k \ldots k \\ q=-p \ldots p}} \xi_{hqt}^{kpr} \, \mathcal{M}_{hq}^{kp} \quad ,
\end{equation}
where the coefficients $\xi_{hqt}^{kpr}$ are essentially Clebsch-Gordan coefficients combining the two irreducible representations corresponding to $k$ and $p$, and $t=-r, \dots , r$. 

For the purpose of this work, we can discard the radial dependence of the multipoles, so that the integral in \pref{eq:spherical-multipole} can be reduced to $\int Y^k_h(\theta,\phi)\ \varrho^{p}_q(\theta,\phi)\ d\Omega$, with a suitably radially averaged density. 
Expressing this density in terms of the usual density matrix with respect to spin and angular momenta then leads to:   
\begin{widetext}
\begin{align}\label{eq:irreducible-multipoles}
    w^{kpr}_t 
    = \sum_{\substack{h=-k \ldots k \\ q=-p \ldots p}} \xi_{hqt}^{kpr}
    \int Y^k_h(\theta,\phi)\ \varrho^{p}_q(\theta,\phi)\ d\Omega     
    = \sum_{ll'}\sum_{\substack{h=-k \ldots k \\ q=-p \ldots p}} \xi_{hqt}^{kpr} 
    \sum_{\substack{s=-\sigma \ldots \sigma \\ s^{\prime}=-\sigma \ldots \sigma}} 
        \langle s| \sigma_q^p | s^{\prime} \rangle 
    \sum_{\substack{m=-l \ldots l \\ m^{\prime}=-l' \ldots l'}} 
        \langle lm| v_h^k | l'm^{\prime} \rangle \rho^{}_{lms,l'm's'}  \quad .
\end{align}
\end{widetext}
Here, $v^k_h$ is the spherical tensor operator corresponding to the spatial degrees of freedom in \pref{eq:spherical-multipole}, emerging from the integral over three spherical harmonics in \pref{eq:irreducible-multipoles} with a suitably chosen normalization~\cite{Bultmark_et_al:2009}. 
Thus, the different indices of $w^{kpr}_t$ refer to the full tensor rank ($r$) and corresponding tensor component ($t$), as well as the underlying spatial/angular tensor rank ($k$) and spin tensor rank ($p$). 
In particular, $p=0, 1$ labels charge and magnetic multipoles, respectively. 

In systems with spatial inversion symmetry, which is the case for all systems considered in this work, only multipoles with even $k$ are nonzero. 
Furthermore, we restrict our analysis to the $l=l'=2$ components of the density matrix centered on the transition metal cations, which make up the largest part of the magnetization density.
This sector of the density matrix only allows for nonzero multipoles with $k = \{ 0, 2, 4 \}$. For the case of the charge multipoles ($p=0$), this corresponds to monopoles, quadrupoles, and hexadecapoles. 
Regarding the magnetic multipoles, we limit our interest to those with maximum rank, $r=k+p$, i.e., dipoles, octupoles, and triakontadipoles. Note that for the case of the magnetic octupole, this highest rank component coincides with the totally symmetric traceless part of the full tensor~\cite{Urru/Spaldin:2022}. 

In the following, to ease notation by avoiding too many indices, we use symbols $\mathcal{M}_t$, $\mathcal{Q}_t$, $\mathcal{O}_t$, $\mathcal{H}_t$, and $\mathcal{T}_t$ for magnetic dipoles, charge quadrupoles, magnetic octupoles, charge hexadecapoles, and magnetic triakontadipoles, respectively. 
Furthermore, for the case of quadrupoles and octupoles, we designate the spatial character by $\left\{ xy, yz, 3z^2-r^2, xz, x^2-y^2 \right\}$ corresponding to $t=-2, \dots, 2$, respectively, noting that for a collinear magnetization density along $z$, the $|t|=3$ components of the octupole are zero~\cite{Urru/Spaldin:2022}.
To give some intuition regarding the anisotropy represented by different multipoles, \pref{fig:multipolescartoon_nl} depicts two examples corresponding to a charge density with $\mathcal{Q}_{x^2-y^2} \neq 0$ and a magnetization density characterized by $\mathcal{O}_{xz}$.

To induce a specific multipole on a given site, we employ the constrained DFT approach proposed in  Ref.~\cite{Schaufelberger_et_al:2023}, which minimizes the total energy functional, $E_0[\bm{\varrho}(\mathbf{r})]$, with an added constraint:
\begin{widetext}
\begin{align}\label{eq:3}
E[\{w^{kpr}_t\}] = \min_{\bm{\varrho}(\mathbf{r}), s^{kpr}_t} \Bigg( E_\text{0}[\bm{\varrho}(\mathbf{r})] - \sum_{kprt} s^{kpr}_t \Big( w^{kpr}_t[\bm{\varrho}(\mathbf{r})] - \tilde{w}^{kpr}_t \Big) \Bigg) \ .
\end{align}
\end{widetext}
Here, $\bm{\varrho}(\mathbf{r})$ represents all components of the generalized density defined in \pref{eq:0}, $\tilde{w}^{kpr}_t$ denotes the specific value to which the $w^{kpr}_t$ multipole is constrained, and $s^{kpr}_t$ is a corresponding Lagrange multiplier. 
In practice, $s^{kpr}_t$ represents the strength of an orbital- and spin-dependent local potential shift that is added to the resulting Kohn-Sham potential in order redistribute the electrons such that they produce the corresponding multipole.

Since in this work we only need to vary the local multipoles without constraining them to a specific fixed value, we do not perform a minimization with respect to $s^{kpr}_t$ and instead vary the strength of the shift-potential, $s^{kpr}_t$, within a certain range and then monitor the resulting multipoles.
For simplicity, and to avoid complications with potentially induced noncollinearities in the magnetization density, we only constrain local charge multipoles, in most cases only the quadrupoles, and we only apply constraints to one specific component at a time.
Thereby, we generally apply constraints of identical strength, but varying signs, to all transition metal cations in the unit cell. We distinguish A-, C-, and G-type patterns with different relative signs on these sites corresponding to wave vectors $\mathbf{q}_\text{A} = \left(0, 0, 0.5\right)$, $\mathbf{q}_\text{C}=\left(0.5, 0.5, 0 \right)$, and $\mathbf{q}_\text{G} = \left(0.5, 0.5, 0.5\right)$ defined in terms of the reciprocal vectors of the underlying pseudocubic lattice.

\subsection{Quantitative measures for the NRSS}
\label{sec:NRSS}

In order to establish a general relation between the magnitude of certain multipoles and the magnitude of the resulting NRSS, we compare three different possible measures for the overall spin splitting, with the intention to identify an ensemble quantity that captures the key characteristics without relying on features that might be specific to only certain bands or the need to manually inspect the band-structure for each individual case.
We note that, while the definition of a reliable integrated measure of the NRSS is very desirable for the purpose of this work, in other cases, e.g., for specific applications in spin transport, the focus is more on achieving a pronounced spin splitting of some bands immediately around the Fermi level. 

To this end, we first define the local, band- and $\mathbf{k}$-resolved NRSS as $\Delta(\nu,\mathbf{k}) = \epsilon_{\nu,\uparrow}(\mathbf{k})-\epsilon_{\nu,\downarrow}(\mathbf{k})$, where $\nu$ is the band index and $\epsilon_{\nu,\sigma}(\mathbf{k})$ are the corresponding band energies with spin projection $\sigma$, and the bands for each $\sigma$ are simply indexed in order of increasing energy. We note that for cases where the local NRSS becomes larger than the typical energy difference between subsequent bands, this simple indexing can obviously lead to inconsistencies between spin-up and spin-down bands corresponding to the same $\nu$, such that the local spin splitting is taken between bands that are not necessarily degenerate in the non-altermagnetic limit. We will come back to this problem later in \pref{sec:SCO-quantify}.
 
We then define a first simple measure for the strength of the overall NRSS as:
\begin{equation}
\Delta_\text{max}=\max_{\nu,\mathbf{k}}|\Delta(\nu,\mathbf{k})| \quad , 
\end{equation}
i.e., the maximum absolute value of the local spin splitting across a certain subset of bands at all $\mathbf{k}$-points.
A second measure is obtained by averaging the local NRSS over a subset of bands and all $\mathbf{k}$-points belonging to a high-symmetry path (hsp) where a strong NRSS is expected to arise:
\begin{equation}
\Delta_\text{avg,hsp} = \frac{1}{N_{\nu}N_\mathbf{k}}\sum_\nu\sum_\mathbf{k}^\text{hsp}|\Delta(\nu,\mathbf{k})| \quad .
\end{equation}
Here, $N_\nu$ and $N_\mathbf{k}$ are the number of bands and $\mathbf{k}$-points, respectively, that are included in the averaging, and we use the absolute value, $|\Delta(\nu,\mathbf{k})|$, to prevent spurious cancellations between bands with opposite signs of the NRSS. 
Note that this measure requires prior knowledge about the $\mathbf{k}$-space distribution of the NRSS. Here, we always use a high-symmetry path along the $\mathbf{k}$-space direction reciprocal to the real space direction of highest charge accumulation or depletion of the relevant quadrupole.
Finally, we define a third measure similar to $\Delta_\text{avg,hsp}$, but now averaged over all $\mathbf{k}$-points throughout the whole Brillouin zone (BZ):
\begin{equation}
\Delta_\text{avg,BZ}=\frac{1}{N_{\nu}N_\mathbf{k}}\sum_\nu\sum_k^\text{BZ}|\Delta(\nu,\mathbf{k})| 
\quad . 
\end{equation}
In principle, comparing $\Delta_\text{avg,BZ}$ and $\Delta_\text{avg,hsp}$ allows to establish how confined the NRSS is in $\mathbf{k}$-space.

\subsection{Computational details}

Within the DFT framework, we perform electronic structure calculations using the plane wave-based projector augmented wave (PAW) method~\cite{Blochl:1994, Kresse/Joubert:1999} implemented in the ``Vienna Ab-Initio Simulation Package'' (VASP)~\cite{Kresse/Hafner:1993, Kresse/Furthmuller:1996}, employing the exchange-correlation functional of Perdew, Burke, and Ernzerhof~\cite{Perdew/Burke/Ernzerhof:1996}. We use the standard PAW potentials included in VASP, with all relevant semi-core states included as valence electrons only for relaxations, a plane-wave energy cutoff set to $700$\,eV, and a convergence threshold for the total energy of $10^{-8}$ eV.

For SrCrO$_3$, we use a unit cell corresponding to a \mbox{$\sqrt{2} \times \sqrt{2} \times 1$} supercell of the primitive cell of the underlying ideal perovskite structure, in order to accommodate the C-type AFM order. Calculations are performed on an $11 \times 11 \times 15$ $\Gamma$-centered $\mathbf{k}$-grid.
The symmetry lowering due to the C-type AFM order leads to a simple tetragonal structure with relaxed lattice parameters of $a=3.86$\,\AA\ and $c/a=0.98$. This structure is then kept fixed for all subsequent calculations presented here.

Our calculations for LaVO$_3$ are based on the experimentally obtained structure with $Pbnm$ symmetry from Ref.~\cite{Bordet_et_al:1993}, using the corresponding primitive unit cell which corresponds to a $\sqrt{2} \times \sqrt{2} \times 2$ supercell of the underlying simple perovskite structure.
We then use AMPLIMODES~\cite{Orobengoa_et_al:2009} to decompose the structural distortion of this $Pbnm$ structure relative to the ideal cubic perovskite structure (with $Pm\bar{3}m$ symmetry) into symmetry-adapted modes, and construct hypothetical structures that contain only one specific symmetry adapted mode with varying amplitude, while keeping the lattice parameters fixed to that of the undistorted $Pm\bar{3}m$ reference structure (with the same volume per formula unit as the experimental $Pbnm$ structure). 
For all these structures, we perform calculations on a $12\times 12 \times 8$ $\Gamma$-centered $\mathbf{k}$-grid.

The required modifications to the VASP code that allow to perform calculations with constrained multipoles are provided by the ``multipyles'' package~\cite{Merkel:2023}, which is also used to perform the subsequent analysis of the resulting multipoles.

\section{Results}

\subsection{Quantifying the induced altermagnetic spin-splitting in SrCrO$_3$}
\label{sec:SCO}

SrCrO$_3$ has been investigated as a rare example for an antiferromagnetic metal~\cite{Carta/Ederer:2022, Carta/Panda/Ederer:2024, Komarek_et_al:2011, Ortega-San-Martin_et_al:2007, Zhang_et_al:2015, Lee/Pickett:2009, Qian_et_al:2011}, and has also been reported to exhibit a tendency towards orbital order in combination with a Jahn-Teller distortion and a metal-insulator transition under tensile epitaxial strain~\cite{Carta/Ederer:2022}. However, the Kugel-Khomskii coupling between orbital and spin-order leads to a preferential G-type AF orbital polarization in combination with C-type AFM order. The resulting symmetry is not compatible with the presence of a ferroically ordered magnetic multipole and the emergence of a NRSS, even though the resulting electronic structure can be classified as ``anti-altermagnetic''~\cite{Meier_et_al:2025}.

Here, we use C-type AFM SrCrO$_3$ as a simple model material, with no spontaneous orbital order. We induce specific multipoles by applying corresponding constraints to the charge density~\cite{Schaufelberger_et_al:2023} and vary their magnitude. This allows us to monitor the resulting NRSS as function of different multipole moments in the absence of any structural distortion, i.e., the symmetry breaking and emerging NRSS in this case is of purely electronic origin.

To ensure a symmetry breaking compatible with a NRSS, we always impose a C-type AF arrangement of local charge multipoles on top of the C-type AFM configuration. This means that the product of the local magnetic dipole and the charge quadrupole (or hexadecapole), which is symmetry-equivalent to the local magnetic octupole (or triakontadipole), has the same sign on every site and thus allows for a ferroic octupolar (or triakontadipolar) order. 

\subsubsection{Induced multipoles}
\label{sec:SCO-multipoles}

\begin{figure*}
    \centering
    \includegraphics[width=1\textwidth]{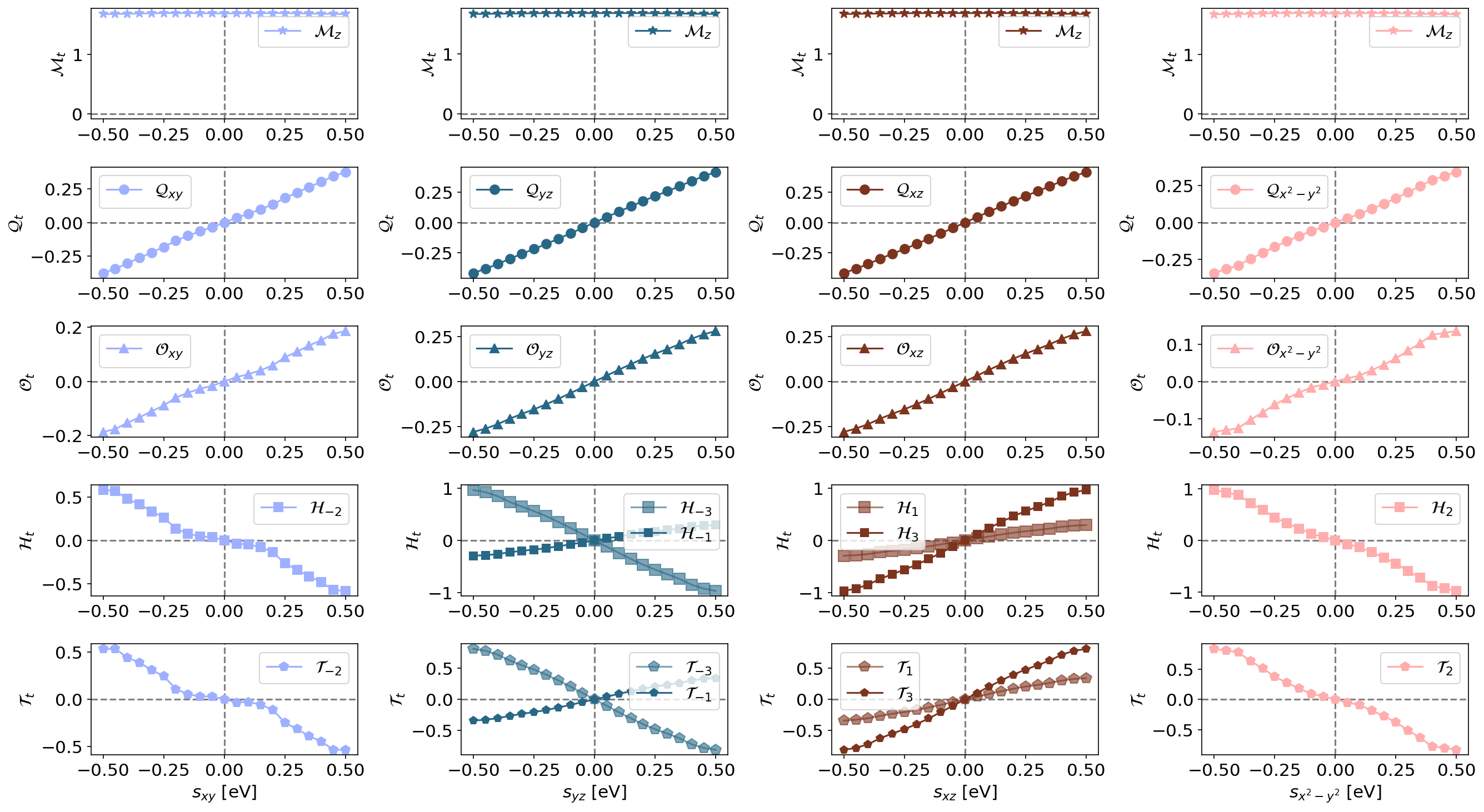}
    \caption{Dependence of charge and magnetic multipoles corresponding to one of the two Cr atoms in the unit cell on an applied charge quadrupolar perturbation. From left to right, the columns correspond to different perturbations: $s^{202}_{-2}=s_{xy}$, $s^{202}_{-1}=s_{yz}$, $s^{202}_{1}=s_{xz}$, $s^{202}_{2}=s_{x^2-y^2}$.}
    \label{fig:combined_ssvs}
\end{figure*}

In \pref{fig:combined_ssvs} we show the magnetic dipole moments, the emerging C-type antiferroically ordered charge multipoles, and the ferroically ordered higher order magnetic multipoles as function of the strength of each specific quadrupolar perturbation applied. The reported values correspond to one of the two Cr atoms in the unit cell. Due to the C-type antiferroic arrangement, the magnetic dipole and charge multipoles on the other, otherwise symmetry-equivalent Cr atom have opposite signs, while the signs of the higher order magnetic multipoles are identical.

One can see that the magnetic dipoles are essentially unaffected by the applied perturbation and remain constant, while all the other multipoles are zero in the unperturbed case and change sign if the sign of the applied local perturbation potential is reversed. The induced charge quadrupoles are linearly related to the perturbation over the entire considered range.
The emergence of certain higher order charge multipoles is consistent with the specific symmetry-breaking introduced by the applied perturbation. 

A nonzero charge multipole then also implies a corresponding nonzero magnetic multipole with the same symmetry as the product of that charge multipole and the magnetic dipole. Thus, as expected, each induced charge quadrupole is accompanied by a ferroically ordered magnetic octupole. Since the magnetic dipoles remain constant, the dependence of the magnetic octupoles on the applied pertubation mirrors, to a good approximation, the dependence of the corresponding charge quadrupole. The same relation can be observed between the charge hexadecapoles and corresponding magnetic triakontadipoles.
However, we also note that the octupoles are not strictly equal to the product of the corresponding charge quadrupole and the magnetic dipole, which can be seen from small deviations from linear behavior specifically for $\mathcal{O}_{xy}$ and $\mathcal{O}_{x^2-y^2}$.
We also note that the induced C-type anti-ferroically aligned charge hexadecapoles do not necessarily show a linear dependence on the applied perturbation, in particular for the case of the $s_{xy}$ and $s_{x^2-y^2}$ perturbations. This behavior is then also mirrored in the corresponding magnetic triakontadipoles.

\subsubsection{Quantification of the (average) NRSS}
\label{sec:SCO-quantify}

\begin{figure*}
    \centering
    \includegraphics[width=1\textwidth]{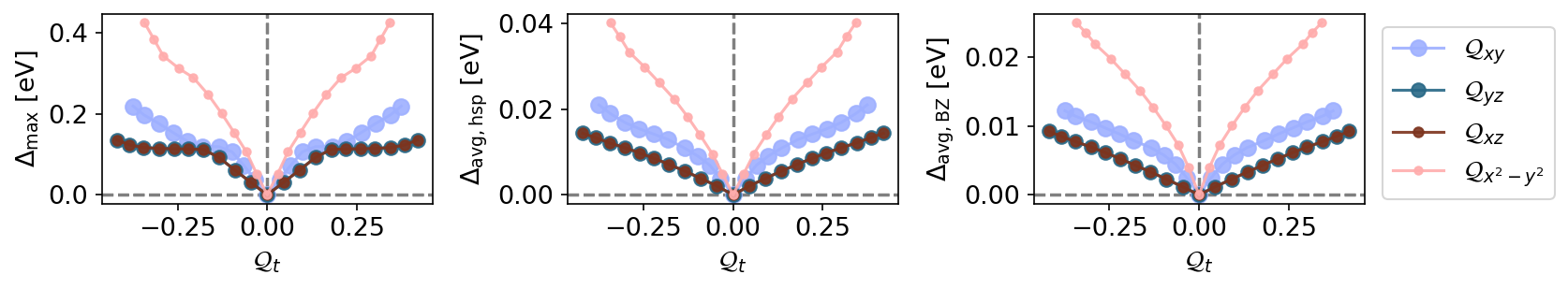}
    \caption{Evolution of the different measures for the overall NRSS as function of different induced local charge quadrupoles. The three panels show the global maximum, $\Delta_\text{max}$ (left), the average along the corresponding ``ideal'' high-symmetry $\mathbf{k}$-paths, $\Delta_\text{avg,hsp}$ (center), and the full BZ average $\Delta_\text{avg,BZ}$ (right).}
    \label{fig:nrssvsquad}
\end{figure*}

In \pref{fig:nrssvsquad} we show how the three different measures for the NRSS, defined in \pref{sec:NRSS}, behave as a function of the different induced charge quadrupoles. 
We first observe that the overall trends are comparable for all three measures and that each type of symmetry breaking has a distinct impact on the NRSS. In particular, the strongest NRSS is obtained from $\mathcal{Q}_{x^2-y^2}$, i.e., when the direction of maximum charge accumulation/depletion is aligned with the Cr-O bonds. The difference between the effect of $\mathcal{Q}_{xy}$ and $\mathcal{Q}_{xz}$/$\mathcal{Q}_{yz}$ results from the different orientations of these quadrupoles  relative to the C-type wavevector, $\mathbf{q}_\text{C} = \left( 0.5, 0.5, 0\right)$.
Moreover, while there is a clear linear dependence of $\Delta_\text{avg, hsp}$ and $\Delta_\text{avg, BZ}$ on both $\mathcal{Q}_{yz}$ and $\mathcal{Q}_{xz}$, the NRSS as function of $\mathcal{Q}_{xy}$ and $\mathcal{Q}_{x^2-y^2}$ shows some deviations from a purely linear behavior (further discussion of this behavior is provided at the end of this subsection).

Regarding the different measures to quantify the NRSS, it is apparent that $\Delta_\text{max}$ exhibits a more complex behavior as function of $Q_t$ compared to $\Delta_\text{avg, hsp}$ and $\Delta_\text{avg, BZ}$. In particular, $\Delta_\text{max}$ suffers from the fact that the underlying $\Delta(\nu, \mathbf{k})$ is obtained from the difference between spin-up and spin-down bands with the same band index $\nu$, ordered simply from lowest to highest energy. Thus, once the local NRSS for a particular band becomes larger than the initial energy difference to the next highest or lowest band with the same spin, the bands can cross, and the assignment between corresponding spin-up and spin-down bands can become incorrect. 

\begin{figure}
    \includegraphics[width=0.45\textwidth]{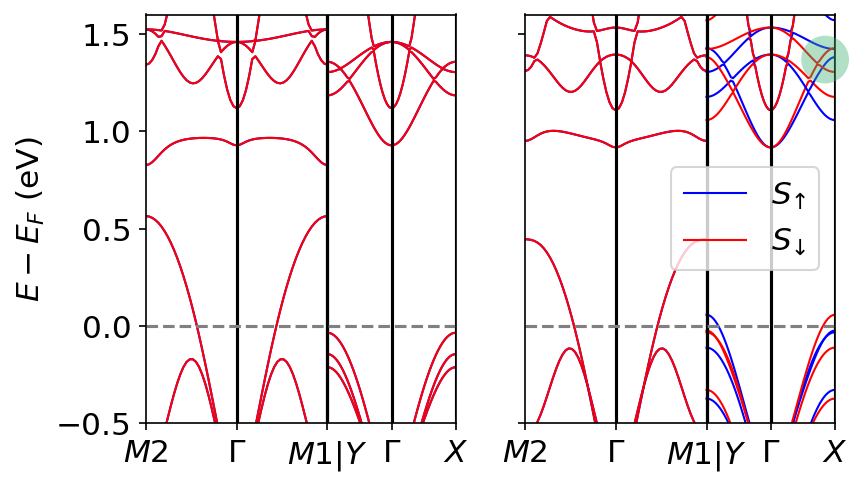}
    \caption{Bandstructure of C-AFM SrCrO$_3$ without (left) and with (right) application of a quadrupolar perturbation of $s_{xy}=0.5$\,eV. 
    The green circular area highlights a region where problematic bands-crossing occur, leading to potentially incorrect assignment between spin-up and spin-down bands and thus a slight underestimation of the total NRSS.}
    \label{fig:scrobs}
\end{figure}

Band crossings between adjacent bands with the same spin character can indeed be recognized in the bandstructures shown in \pref{fig:scrobs}, which correspond to the cases without and with an applied $s_{xy}$ perturbation. While the bandstructure for the unperturbed case does not show any spin-splitting, a NRSS appears for nonzero $s_{xy}$ along the path $Y-\Gamma-X$, while the bands along the $\Gamma-M$ directions remain spin-degenerate. This is consistent with the symmetry of the $\mathcal{Q}_{xy}$ quadrupole, noting that the high symmetry $\mathbf{k}$-points in \pref{fig:scrobs} are indexed according to the simple tetragonal BZ, while the local $x$-$y$ axes are rotated by 45$^\circ$ relative to the basal plane lattice vectors. 
Importantly, we highlight in green one of the problematic band crossings, arising between the bands evolving along the $\Gamma–X$ direction from the first and the fourth band above the Fermi energy at $\Gamma$, which split in opposite ways. For the former, the down-spin state is higher in energy, whereas for the latter, the up-spin state is higher. As a result, $\Delta(\nu,\mathbf{k})$ couples the two bands incorrectly for $\mathbf{k}$-vectors close to $X$, leading to a small underestimation of the overall NRSS.

Such bands-crossings are more likely to occur for larger spin splitting, preventing the calculated $\Delta_\text{max}$ from capturing the actual maximum splitting.
The other two definitions of $\Delta$ of course suffer from the same problem, but the averaging over bands and $\mathbf{k}$-points seems to provide an effective way to partially suppress this issue. 
Including more $\mathbf{k}$-points in the averaging can thus be expected to further mitigate this band-crossing problem even at larger perturbations, in particular if the larger set of $\mathbf{k}$-points includes regions in the BZ where the spin splitting is weaker. This is consistent with the slightly more regular behavior of $\Delta_\text{avg, BZ}$ compared to $\Delta_\text{avg, hsp}$ in \pref{fig:nrssvsquad}. 
Nevertheless it is important to mostly focus on the low perturbation regime, where bands-crossings are less likely to occur.

With this in mind, we can now have a closer look at the dependence of $\Delta_\text{avg, BZ}$ on $\mathcal{Q}_{xy}$ and $\mathcal{Q}_{x^2-y^2}$ in \pref{fig:nrssvsquad}. As noted previously, both cases show deviations from perfectly linear behavior. For increasing $|\mathcal{Q}_{x^2-y^2}|$, one can first recognize a gradual deviation from linear behavior, typical for the increasing influence of higher order terms, and then observe a kink-like feature around $\mathcal{Q}_{x^2-y^2} = 0.3$ (corresponding to $s_{x^2-y^2} = 0.4$\,eV). The NRSS as functions of $\mathcal{Q}_{xy}$ also exhibits as somewhat sudden change in slope around $\mathcal{Q}_{xy} = 0.1$ (corresponding to $s_{xy} = 0.15$\,eV), separating two essentially linear regimes below and above.
Inspection of the induced multipole moments shown in \pref{fig:combined_ssvs}, suggests that these more sudden features in the NRSS appear to be related to corresponding features in the higher order  multipoles, such as the emergence of a more substantial $\mathcal{H}_{-2}$ for $s_{xy} > 0.15$\,eV, and a noticeable saturation of $\mathcal{H}_2$ (but also $\mathcal{O}_{x^2-y^2}$) for $s_{x^2-y^2} > 0.4$\,eV. 
This indicates that the NRSS might not necessarily be determined exclusively by the lowest order nonzero multipole, but that higher order multipoles can also contribute. We will come back to this point in \pref{sec:lvo}.

\subsubsection{Band- and $\mathbf{k}$-resolved NRSS}

\begin{figure}
    \includegraphics[width=0.5\textwidth]{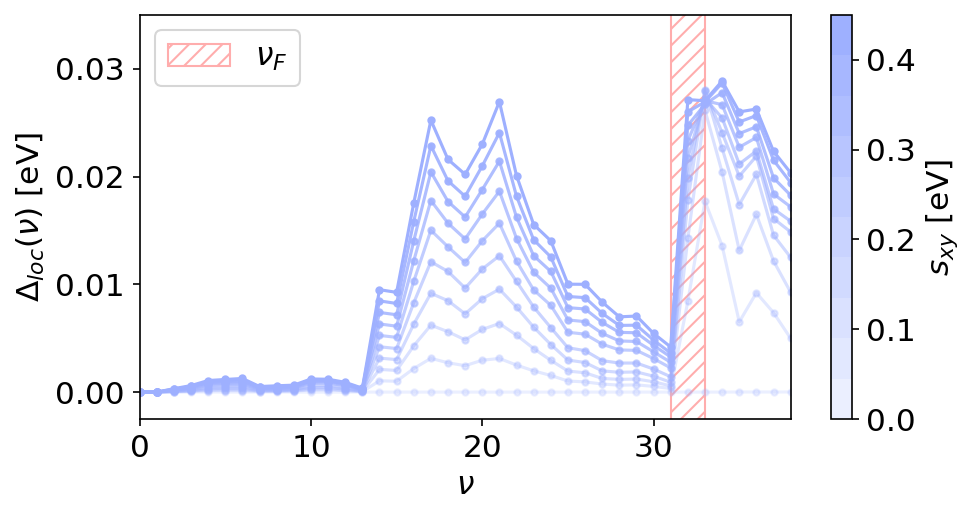}
    \caption{Evolution and distribution of the NRSS across band indices in SrCrO$_3$ for increasing quadrupolar perturbation $s_{xy}$. For reference, we indicate all bands that cross the Fermi energy at least at one $\mathbf{k}$-point in the BZ by $\nu_F$ .}
    \label{fig:nrssvsbandidx}
\end{figure}

\begin{figure}
    \includegraphics[width=0.35\textwidth]{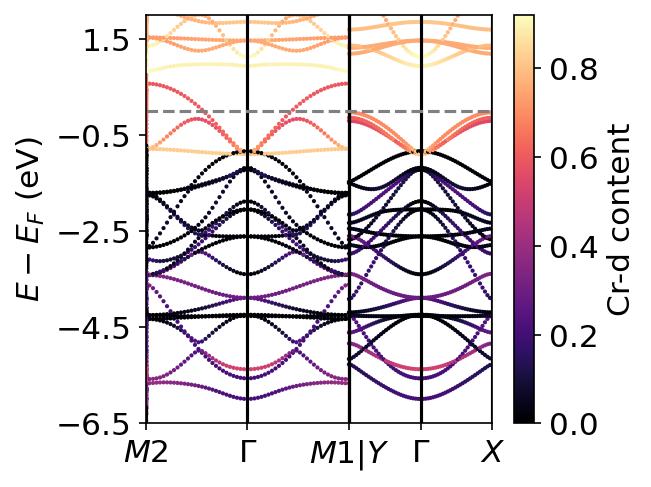}
    \caption{Unperturbed bandstructure of C-AFM SrCrO$_3$ projected onto the Cr $d$-orbitals. The projection highlights two distinct regions where the Cr $d$ contributions is significant: around and above the Fermi level, where the bands exhibits dominant Cr $d$ character, and towards the bottom of the otherwise O $p$-dominated bands, which also exhibit strong hybridization with the Cr $d$ states.}
    \label{fig:pjbandsco}
\end{figure}

Next, to further analyze the spin splitting resulting from the different induced multipoles, we also look at the energy-resolved (or rather band-resolved) and also the $\mathbf{k}$-resolved NRSS.
In \pref{fig:nrssvsbandidx} we show the BZ-averaged NRSS calculated for each band, defined as $\Delta_\text{loc}(\nu)=\frac{1}{N_{\mathbf{k}}}\sum_\mathbf{k}\Delta(\nu,\mathbf{k})$.
It can be seen that the NRSS predominantly involves two groups of bands, one corresponding to energies around or above the Fermi level, and one corresponding to energies of some eV below the Fermi level. Inspection of the orbitally-projected band structure shown in \pref{fig:pjbandsco} indicates that the strong spin-splitting of these bands correlates with the corresponding Cr $d$ character, with the groups at higher and lower energies corresponding to the anti-bonding and bonding bands formed from the hybridization between atomic Cr $d$ and O $p$ states, respectively.
That the spin-splitting correlates with the Cr $d$ character is of course expected, since these states carry most of the magnetization density. 

From \pref{fig:nrssvsbandidx} one can also see that for small perturbations, $s_{xy}$, the largest spin-splitting arises in the Cr $d$-dominated antibonding bands around and above the Fermi level, but that the splitting appears to saturate for larger perturbations. This is likely due to increased presence of band-crossings for stronger perturbations, which can then shadow a further increase of the NRSS. 
In contrast, the bands at lower energies with weaker Cr $d$ character show a more regular, gradual increase of the NRSS with increasing perturbation.
Thus including these bands in the definition of $\Delta_\text{avg, BZ}$ also dampens the unwanted effect of potential band crossings and results in a more faithful representation of the overall average NRSS in the system.

\begin{figure}
    \includegraphics[width=0.45\textwidth]{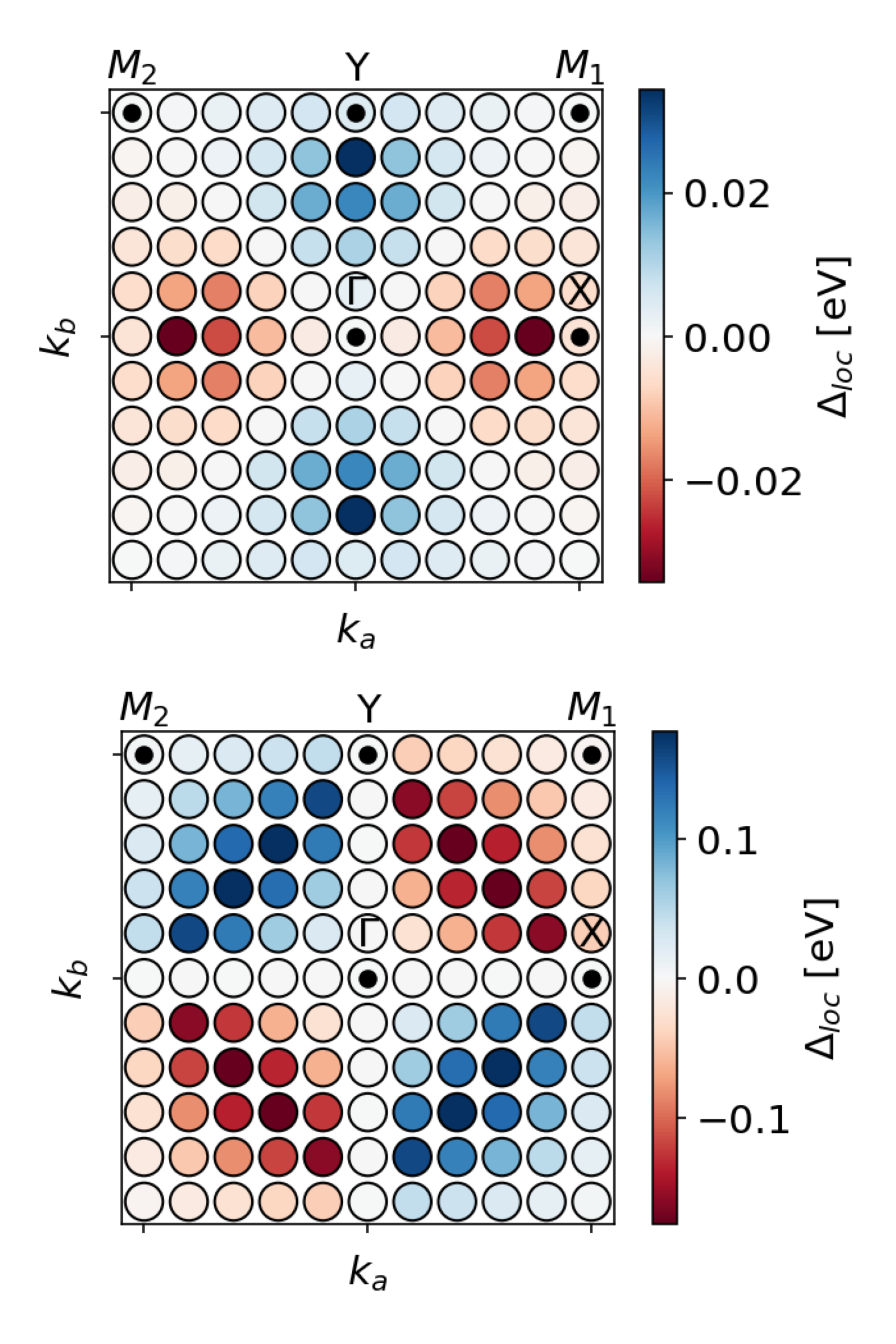}
    \caption{$\mathbf{k}$ dependence of the band-averaged NRSS, $\Delta_\text{loc}(\mathbf{k})$ for different quadrupolar perturbations, plotted for $k_z=0$. The upper panel corresponds to a $s_{xy}=0.5$\,eV, while the lower panel corresponds to $s_{x^2-y^2}=0.5$\,eV. Note that the basal-plane reciprocal lattice directions $k_a$ and $k_b$ are rotated by 45$^\circ$ around $z$ relative to the cartesian axes.}
    \label{fig:multipolescartoon} 
\end{figure}

\pref{fig:scrobs} already indicates to some extent how the symmetry of the induced multipole is reflected in the $\mathbf{k}$-dependence of the resulting NRSS for the case of a $\mathcal{O}_{xy}$ octupole, which translates into a $\mathbf{k}$- and spin-dependence of the form $\sigma_zk_xk_y$~\cite{Bhowal/Spaldin:2024}. 
Consequently, a clear NRSS is expected along the $[110]$-type directions in $\mathbf{k}$-space, with opposite signs for $[110]$ relative to $[\bar{1}10]$, while spin degeneracy is expected to be preserved along the $[100]$ and $[010]$ directions \cite{Smejkal/Sinova/Jungwirth:2022, Urru_et_al:2025}. This is indeed the case in the perturbed bandstructure shown in \pref{fig:scrobs}, noting again that in our setup the tetragonal basal plane lattice vectors (and thus the corresponding reciprocal lattice vectors) are rotated by 45$^\circ$ around the $z$-axis relative to the $x$-$y$ axes that are aligned along the Cr-O bonds.

To also obtain a quantitative idea of the $\mathbf{k}$-resolved NRSS, we now analyze the local NRSS averaged over all bands but for fixed $\mathbf{k}$,
$\Delta_\text{loc}(\mathbf{k})=\frac{1}{N_{\nu}}\sum_\nu\Delta(\nu,\mathbf{k})$.
The upper and lower panels of \pref{fig:multipolescartoon} depict $\Delta_\text{loc}(\mathbf{k})$ within the $k_x$-$k_y$ plane (and $k_z=0$), calculated for perturbations $s_{xy}=0.5$\,eV and $s_{x^2-y^2}=0.5$\,eV, respectively. As expected, the sign and strength of the NRSS indeed reflect the reduced spin-space-group symmetry of the system, as described in \cite{Smejkal/Sinova/Jungwirth:2022, Urru_et_al:2025}. Consequently, the NRSS is zero along the nodal planes of the imposed quadrupole, while the strongest spin splitting can be observed vaguely in the regions corresponding to highest charge accumulation or depletion of the correspondong quadrupole in real space. 
Note that here we defined $\Delta_\text{loc}(\mathbf{k})$ without taking the absolute value of $\Delta(\nu,\mathbf{k})$, and in \pref{fig:multipolescartoon} we can indeed discriminate regions with net positive and net negative NRSS, even though the sign of $\Delta(\nu,\mathbf{k})$ is band-dependent. This means that in principle $\Delta_\text{loc}(\mathbf{k})$ could become zero also away from the symmetry-imposed nodal surfaces, and is therefore not necessarily guaranteed to always provide a faithful representation of the $\mathbf{k}$-dependence of the overall NRSS. However, in the present case it appears to be instructive.

\begin{figure}
    \includegraphics[width=0.45\textwidth]{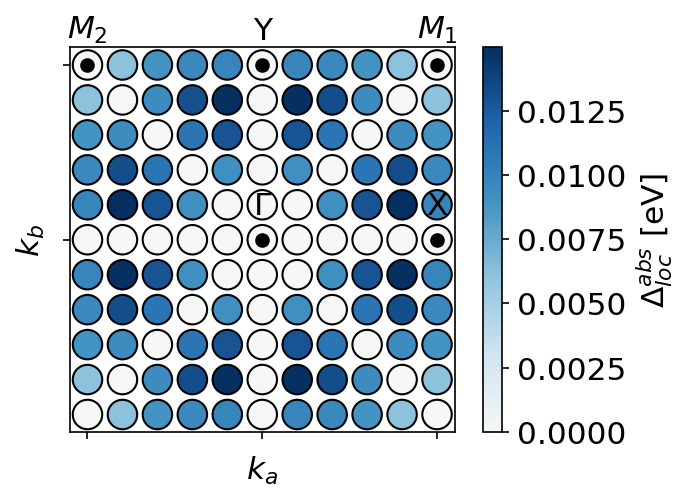}
    \caption{$k$-dependence (for $k_z=0$) of the band-averaged NRSS (here defined in terms of the absolute value of $\Delta(\nu,\mathbf{k})$) for a hexadecapolar perturbation, $s^{404}_{-4}=0.5$\,eV.}
    \label{fig:nrssmaphexa}
\end{figure}

To further test the correspondence between the $\mathbf{k}$-space symmetry of the NRSS and the lowest order nonzero multipole, we now perform a calculation where we induce a charge hexadecapole instead of a charge quadrupole. Specifically we apply a shift-potential of the form $s^{404}_{-4}$, corresponding to a C-type AF ordered $\mathcal{H}_{-4}$ hexadecapole, which is of the form $\mathcal{H}_{-4} \propto  \int \frac{x y\left(x^2-y^2\right)}{r^4} \rho(\mathbf{r})\ d\mathbf{r}$.
The resulting symmetry lowering does not allow for charge quadrupoles or magnetic octupoles, but also induces a ferroically ordered $\mathcal{T}_{-4}$ magnetic triakontadipole.

If one considers the nodal structure of the NRSS produced by the $s^{404}_{-4}$ perturbation shown in \pref{fig:nrssmaphexa}, one recognizes a clear $g$-wave pattern, with nodal planes corresponding exactly to the nodal planes of $\mathcal{H}_{-4}$ in real space, i.e., $k_x=0$, $k_y=0$, and $k_x=\pm k_y$.
Note that here we defined the NRSS as $\Delta^\text{abs}_\text{loc}(\mathbf{k})=\frac{1}{N_{\nu}}\sum_\nu|\Delta(\nu,\mathbf{k})|$ to avoid sign cancellations between different bands, which would indeed prevent a proper characterization of the NRSS in this case.

\subsection{Quantifying the altermagnetic spin-splitting in LaVO$_3$}
\label{sec:lvo}

\begin{figure*}
    \includegraphics[width=0.95\textwidth]{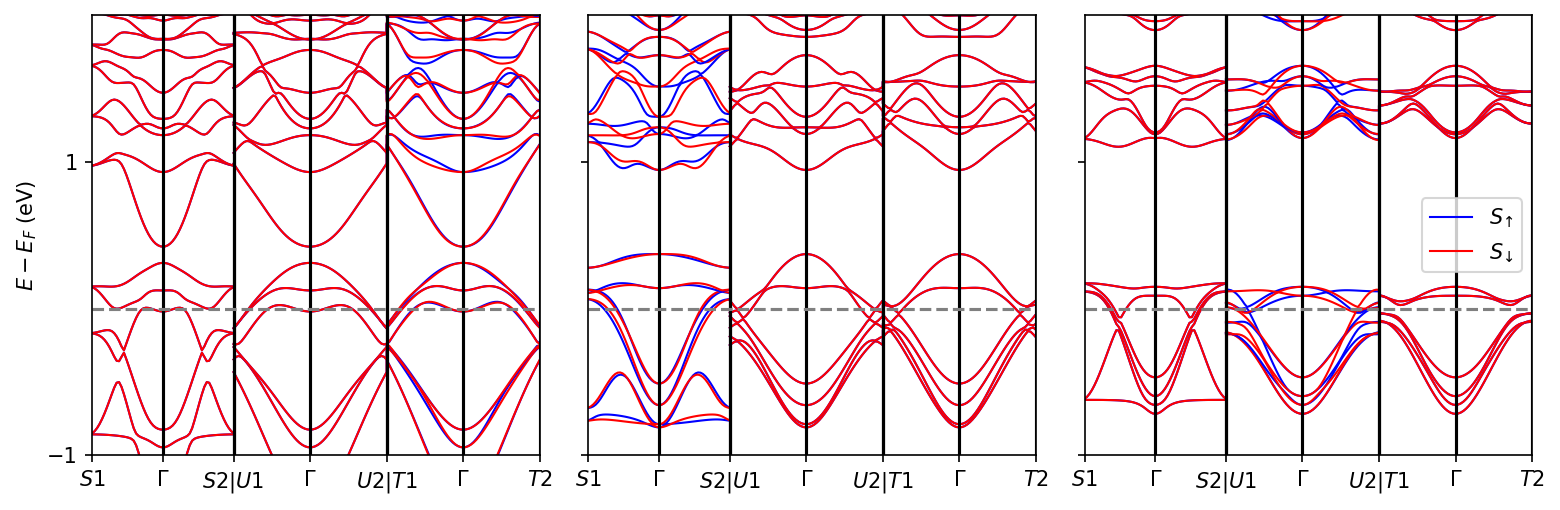}
    \caption{Spin-polarized bandstructures of LaVO$_3$ calculated for A-type AFM (left), C-type AFM (middle), and G-type AFM (right) orders.}
    \label{fig:bslvofd}
\end{figure*}

\begin{figure}
    \includegraphics[width=0.5\textwidth]{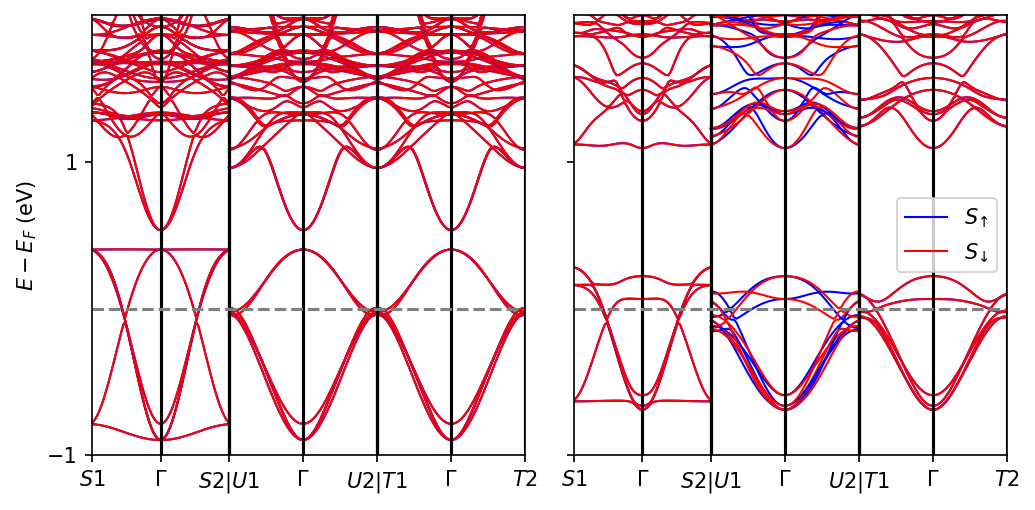}
    \caption{Bandstructure of G-AFM LaVO$_3$ in the parent cubic structure (left) and a distorted structure with only the $R_5^-$ mode included (right).}
    \label{fig:bslvor5}
\end{figure}

The example of SrCrO$_3$ discussed in \pref{sec:SCO} allowed us to explore the relation between multipoles of the electronic charge and magnetization density and the resulting NRSS in a simple and well-defined setting. 
Notably, the corresponding NRSS is caused by a purely electronic symmetry breaking, without any resulting structural distortions, confirming the possibility of purely electronically driven altermagnetism. 

However, since the emergence of these multipoles is not energetically favorable in SrCrO$_3$, they had to be induced artificially via an externally applied constraint. 
Higher order local multipoles are stabilized in certain altermagnets simply by the intrinsic crystal symmetry, such as, e.g., in the rutile structure~\cite{Bhowal/Spaldin:2021}. However, the resulting multipoles and thus the corresponding NRSS cannot easily be tuned.
Another mechanism that can in principle stabilize potentially tunable multipoles is orbital order, and the accompanying Jahn-Teller distortion. The possibility of orbital-order-induced NRSS has been suggested in Ref.~\cite{Leeb_et_al:2024}. However, the Goodenough-Kanamori rules tend to favor combinations of magnetic dipolar and charge quadrupolar order that are incompatible with altermagnetism~\cite{Meier_et_al:2025}.

An alternative way to obtain stable (and potentially tunable) local quadrupole moments is thus through structural distortions that are driven by other mechanisms, such as, e.g., the collective octahedral rotations frequently occurring in perovskite systems.
Altermagnetic spin splittings have indeed been reported based on electronic structure calculations for several distorted perovskites with $Pbnm$ space group symmetry~\cite{Bandyopadhyay/Picozzi/Bhowal:2025, Smejkal/Sinova/Jungwirth:2022a, Yuan/Zunger:2023, Naka/Motome/Seo:2025, Rooj/Saxena/Ganguli:2025, Fernandes_et_al:2024}, even several years before the emergence of the term altermagnetism~\cite{Okugawa_et_al:2018}.
A systematic classification of higher order multipoles and their relation to different distortion modes has recently been presented in Ref.~\cite{Bandyopadhyay/Picozzi/Bhowal:2025}, using LaMnO$_3$ as example.
In this section, we built on this work and establish a quantitative relation between different distortion modes present in $Pbnm$ perovskites, the resulting multipole moments, and the corresponding NRSS. 

We use LaVO$_3$ as an example model material which crystallizes in the $Pbnm$ distorted perovskite structure. We note that below $\sim140$\,K LaVO$_3$ develops additional structural distortion related to orbital order~\cite{Bordet_et_al:1993, Miyasaka_et_al:2003}. However, since in the present case we want to focus exclusively on the effects of the octahedral rotations, we essentially ``switch off'' the tendency for orbital order in LaVO$_3$ by \emph{not} applying a Hubbard correction in our DFT calculations. 

As discussed in \cite{Bandyopadhyay/Picozzi/Bhowal:2025}, the $Pbnm$ distortion allows for local charge quadrupoles, with different quadrupoles arranged in A-, C-, and G-type patterns, respectively. Thus, each quadrupolar component can couple to a matching A, C, or G-type magnetic dipolar order, in all cases resulting in a symmetry that is compliant with a ferroic order of a specific magnetic octupole component. In other words, this means that any $Pbnm$ distorted perovskite with an A-, C,- or G-type AFM order exhibits a spin-group symmetry that is compatible with a NRSS along specific $\mathbf{k}$-space directions.

We first demonstrate that this conclusion can indeed also be confirmed for the case of LaVO$_3$, by imposing each of the three magnetic dipolar order patterns. \pref{fig:bslvofd} shows the calculated band-structures along the relevant $\mathbf{k}$-space directions. Consistent with Ref.~\cite{Bandyopadhyay/Picozzi/Bhowal:2025}, the $Pbnm$ structure allows for an A-type ordered $\mathcal{Q}_{yz}$, a C-type ordered $\mathcal{Q}_{xy}$, and a G-type ordered $\mathcal{Q}_{xz}$ local quadrupole on the V sites. Thus, for each corresponding AFM dipolar order, a NRSS emerges along the [011] ($\Gamma$-T), [110] ($\Gamma$-S), and [101] ($\Gamma$-U) directions in $\mathbf{k}$-space, respectively.

Next, we use the mode decomposition of the experimental structure to isolate the three main distortion modes, $R_5^-$, $M_2^+$, and $X_5^-$, which describe out-of-phase octahedral rotations around [110], in-phase octahedral rotations around $z$, and antipolar displacements of the La cations perpendicular to $z$ (see, e.g., Ref.~\cite{Bandyopadhyay/Picozzi/Bhowal:2025} for a more detailed discussion of the different modes present in $Pbmm$ distorted perovskites). Each of these modes alone already produces a modulation of the charge density corresponding to a G-type, C-type, and A-type wave-vector, respectively. However, while the lowest order emerging local multipole for the $R_5^-$ and $X_5^-$ modes is a quadrupole, the $M_2^+$ mode only results in a local charge hexadacapole while all quadrupolar components remain zero~\cite{Bandyopadhyay/Picozzi/Bhowal:2025}.
Leveraging this mode decomposition, we create distorted structures where only one of these modes appear at a time, while all other modes remain zero, thereby selectively inducing only the specific multipoles to which the selected mode couples. We then tune the amplitude of the corresponding mode and monitor the magnitude of the emerging multipoles as well as the corresponding NRSS. In these calculations, we always impose a matching pattern on the magnetic dipoles, i.e., A-, G, and C-type AFM for the $X_5^-$, $R_5^-$, and $M_2^+$ mode, respectively.

\pref{fig:bslvor5} compares the bandstructures of LaVO$_3$ calculated for the completely undistorted cubic structure (with $Pm\bar{3}m$ symmetry) and a structure with only the $R_5^-$ mode included, with an amplitude corresponding to that of the fully distorted structure, and G-type AFM order. As expected, the undistorted structure does not exhibit any NRSS, whereas the NRSS emerging in the $R_5^-$ distorted structure closely resembles that of the rightmost panel of \pref{fig:bslvofd} for the fully distorted G-AFM structure. While there are some differences, since the presence of the other modes affects the overall bandwidth and breaks additional symmetries, the strong similarity between these two cases suggests that the NRSS in the fully distorted G-AFM structure is indeed dominated mostly by the $R_5^-$ mode.

\begin{figure*}
    \includegraphics[width=1\textwidth]{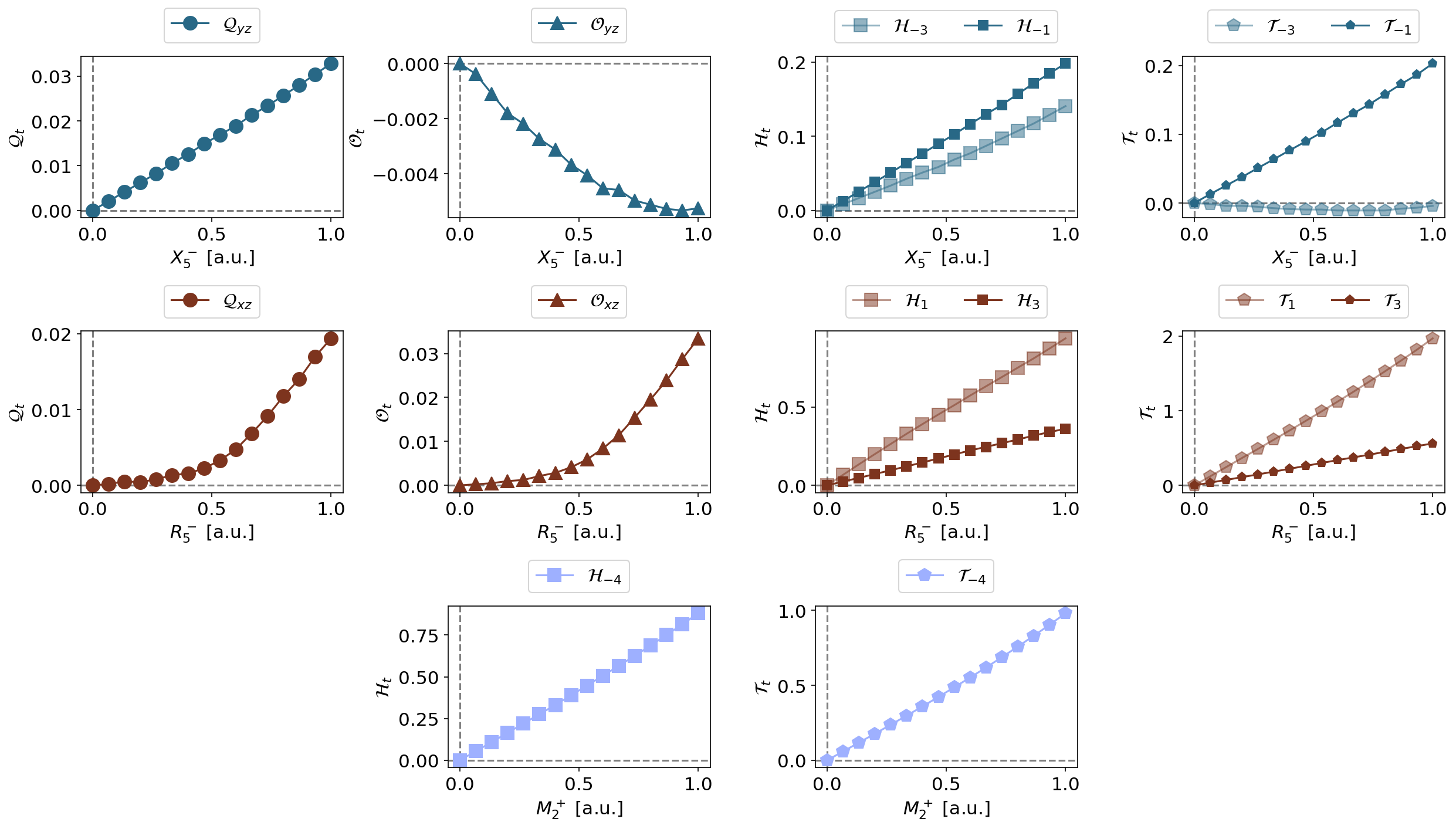}
    \caption{Different local multipole moments (corresponding to one of the V atoms in the cell) induced by the individual distortion modes in LaVO$_3$. We show only the nonzero multipoles that are relevant for the altermagnetic symmetry breaking, i.e., the AF ordered charge multipoles matching the AF pattern of the applied distortion mode and the ferroically ordered magnetic multipoles. Note that the $M_2^+$ mode does note create any nonzero charge quadrupoles or magnetic octupoles.}
    \label{fig:multivsdist}
\end{figure*}

In \pref{fig:multivsdist} we show the dependence of all relevant charge and magnetic multipoles (corresponding to one of the V atoms) as a function of the different distortion mode amplitudes, where in each case only the mode indicated on the $x$-axis of the corresponding plot is nonzero, and an amplitude of 1.0 corresponds to the amplitude of that mode in the experimental structure.

The $X_5^-$ mode (see top row of \pref{fig:multivsdist}) induces an A-type AF order of $\mathcal{Q}_{yz}$, $\mathcal{H}_{-3}$, and $\mathcal{H}_{-1}$ charge multipoles, which all increase linearly with the $X_5^-$ mode amplitude. We  note that the corresponding magnitudes are significantly smaller, by approximately one order of magnitude, compared to the charge multipoles that we induced in SrCrO$_3$ under the quadrupolar constraint. 
The magnitudes of the ferroically ordered magnetic multipoles are even smaller relative to the SrCrO$_3$ case, with $\mathcal{O}_{yz}$ and $\mathcal{T}_{-1}$ about two orders of magnitude smaller, and $\mathcal{T}_{-3}$ even three orders of magnitude smaller. Furthermore, both $\mathcal{O}_{yz}$ and $\mathcal{T}_{-3}$ already show clear deviations from a purely linear dependence on the mode amplitude in the given range. 
Curiously, the negative sign of $\mathcal{O}_{yz}$ is opposite to that of the product of the local magnetic dipole (not shown) and the charge quadrupole. Furthermore, the local magnetic dipole (not shown) is essentially constant as function of the $X_5^-$ mode amplitude. Together with the pronounced deviation from linear behavior of the octupole, this again demonstrates that the magnetic octupole is not merely a product of the magnetic dipole and the charge quadrupole. 
Similar to what has already been briefly alluded to in \pref{sec:SCO-multipoles}, this finding suggests that $\mathcal{O}_t$ and $\mathcal{Q}_t \cdot \mathcal{M}_z$ should in principle be treated as distinct order parameters.

For the $R_5^-$ mode (middle row of \pref{fig:multivsdist}), the dependence of the emerging G-type local quadrupole, $\mathcal{Q}_{xz}$, and ferroically ordered octupole, $\mathcal{O}_{xz}$, show a clear nonlinear dependence on the distortion mode amplitude. Further analysis shows that the G-type $\mathcal{Q}_{xz}$ order indeed corresponds to a different irreducible representation as the $R_5^-$ mode, and therefore the two do not couple linear to each other.
On the other hand, the induced charge hexadecapoles and magnetic triakontadipoles scale linearly with the distortion amplitude. The response in $\mathcal{T}_{1}$ is particularly strong, being one order of magnitude larger than what we induced in SrCrO$_3$. 

Finally, the $M_2^+$ mode leads to the emergence of a linearly increasing $\mathcal{H}_{-4}$ with C-type pattern and of a ferroically ordered $\mathcal{T}_{-4}$. As expected from symmetry, no lower-order multipoles are induced. The magnitudes in this case are comparable to those induced in SrCrO$_3$. 

Comparing the effects of the different modes shown in \pref{fig:multivsdist}, further reveals that the $X_5^-$ mode induces relatively small multipoles compared to the other two modes. This means that the $X_5^-$ mode induces only a relatively weak anisotropy of the charge and magnetization densities around the V cations, which can be attributed to the nature of the $X_5^-$ distortion, which primarily displaces the La atoms, while the other two modes affect the oxygen octahedral environment around the V sites more directly.

\begin{figure*}
    \includegraphics[width=1\textwidth]{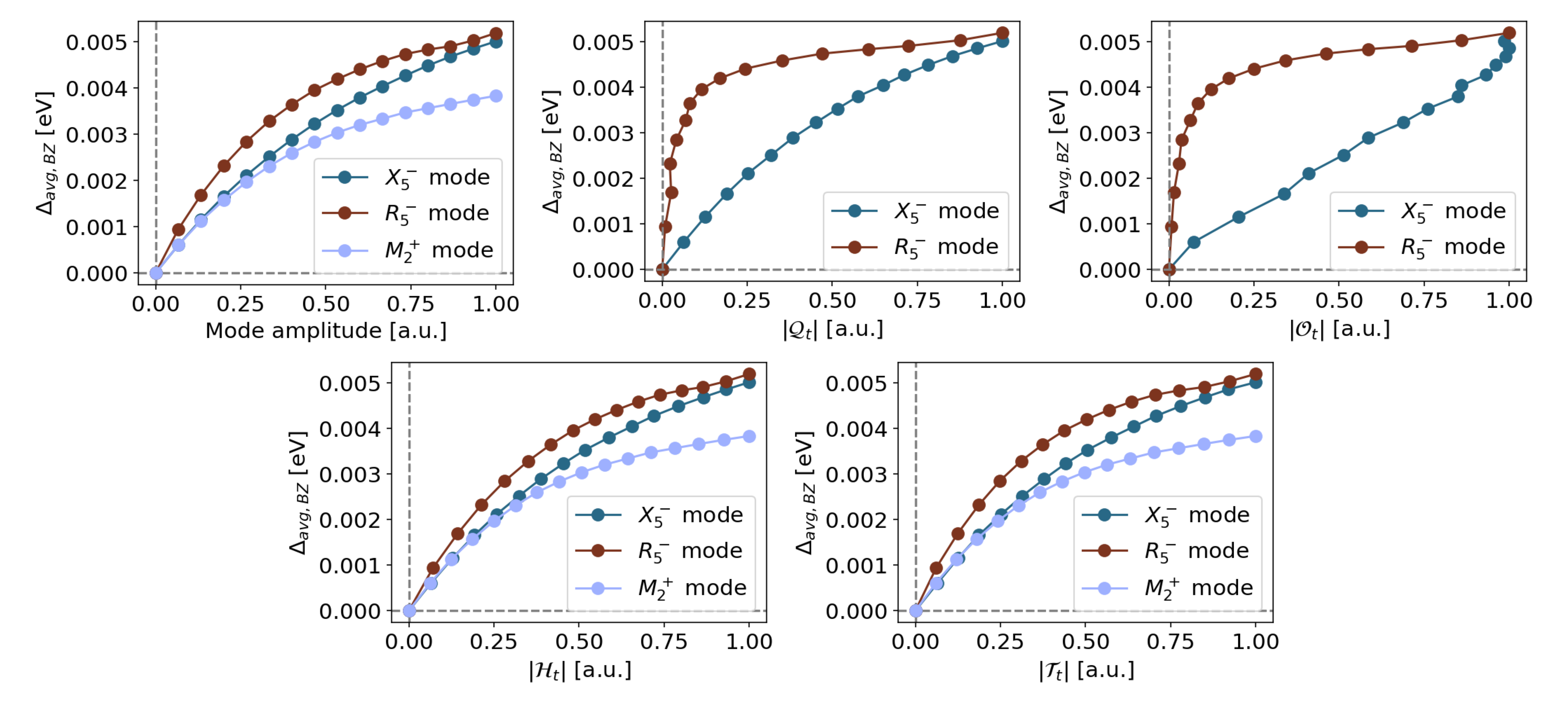}
    \caption{BZ-averaged NRSS, $\Delta_\text{avg,BZ}$, as a function of the three different mode amplitudes and the corresponding induced multipole moments, i.e, the AF aligned charge quadrupole, ferroic magnetic octupole, AF aligned hexadecapole, and ferroic trikontadiapole. In cases where multiple components of the same rank coexist, only the dependence on the dominant contribution (largest magnitude) is plotted. For better comparison across the different modes, all multipoles are normalized relative to their maximum values.}
    \label{fig:nrssvsall_lvo}
\end{figure*}

In \pref{fig:nrssvsall_lvo}, we plot the BZ-averaged NRSS, $\Delta_\text{avg,BZ}$, as function of the three different distortion modes, and also as function of all the various local charge and magnetic multipoles that are induced by these distortion modes. As a first observation, one can note that all three modes give rise to an overall NRSS of comparable magnitude, in spite of the fact that the absolute atomic displacements related to these modes are different (and of course the resulting $\mathbf{k}$-space distribution of the NRSS is different for each mode). We can also note that although some of the multipoles in this case are significantly smaller than what we induced in SrCrO$_3$, the resulting overall spin splitting is only about half of what we discussed in that case.

Focusing on each mode individually, one can see that for the case of the $X_5^-$ mode, the NRSS exhibits a fairly regular behavior, with a linear initial increase as function of all potential order parameters. In particular, the scaling of the NRSS with the lowest order charge and magnetic multipoles, i.e., the charge quadrupole and magnetic octupole, appears consistent with these quantities acting as order parameters for the NRSS, even though the same could also be stated about the higher order charge hexadecapoles and magnetic triakontadipoles.

For the NRSS related to the $R_5^-$ mode, the case is clearly different. While the NRSS exhibits a regular behavior as function of the $R_5^-$ mode amplitude, there is no clear quantitative correlation with the charge quadrupole and magnetic octupole. If plotted as a function of these multipoles, the NRSS first increases steeply, while the quadrupole and magnetic octupole remain nearly zero, and then the NRSS stays nearly constant whereas the two multipoles increase strongly. In contrast, the NRSS scales very regularly both with the hexadecapole and the triakontadipole.
This strongly suggests that in this case the main contribution to the NRSS comes from these higher order multipoles rather than from the lowest order nonzero charge quadrupoles and magnetic octupoles.

Regarding the $M_2^+$ mode, the NRSS also exhibits a regular scaling with these higher order multipoles, noting again that no charge quadrupoles or magnetic octupoles are symmetry-allowed in this case. Nevertheless the corresponding NRSS is nearly of the same magnitude as in the other two cases, in spite of the lacking contributions from these lower-order multipoles.

Collectively, this analysis strongly suggests, that the total NRSS in fact emerges as a superposition of contributions from multipoles with different rank, and is not necessarily dominated quantitatively by the lowest order nonzero multipole. In addition, the simultaneous presence of all three distortion modes (plus two additional minor ones) in the actual $Pbnm$ crystal structure leads to a further superposition of different multipole components with the same rank but different character $t$.
We thus propose that a potential altermagnetic order parameter in general needs to be considered as a multi-dimensional order parameter, with potentially rather high dimension. 

The superposition of the total NRSS stemming from different multipolar components can schematically be expressed as:
\begin{equation}\label{eq:nrss}
    \Delta = \sum_{kprt}\Delta^{kpr}_{t}=\sum_{kprt} \delta^{kpr}_{t} w^{kpr}_{t} \quad ,
\end{equation}
with materials-specific coupling coefficients, $\delta^{kpr}_{t}$, for each relevant multipole.
We note that such a superposition is similar in spirit to a recently proposed decomposition of the $\mathbf{k}$-dependent NRSS in terms of symmetry-adapted plane waves~\cite{Urru_et_al:2025}, even though a potential relation between this symmetry-adapted plane wave decomposition and the decomposition in terms of spherical multipoles in \pref{eq:nrss} still needs to be established.

\section{Summary and Conclusions}

In this work, we have addressed the quantitative relation between the altermagnetic spin splitting and certain multipoles of the charge and magnetization density as corresponding order parameters.
We used SrCrO$_3$ and LaVO$_3$ as model materials for our calculations, but our general conclusions should be applicable to a wide variety of other materials.

We showed that a purely electronic symmetry breaking is sufficient to induce a substantial NRSS, which indicates that the definition of an electronic altermagnetic order parameter is indeed reasonable.
A corresponding symmetry-breaking is generally related to a ferroic order of magnetic multipoles of higher order than the simple dipoles (or, equivalently, to a ferroic order of the product of the local magnetic dipoles and a charge multipole).

For the case of SrCrO$_3$, for which we selectively induced different charge multipoles with a suitable ordering pattern that also allows for a ferroic net magnetic multipole, we find that there is indeed a clear quantitative relation between the emerging NRSS and the corresponding multipole. However, we also found first indications that not only the lowest order nonzero multipole determines the overall strength of the NRSS, but that higher order multipoles also contribute. Nevertheless, the nodal structure of the NRSS in $\mathbf{k}$-space is determined by the lowest order nonzero multipoles, with ferroic octupolar or triakontadipolar order resulting in $d$-wave and $g$-wave symmetry, respectively.

In this context, we also assessed different simple measures to quantify the overall NRSS, and found the BZ-averaged measure most suitable and rather insensitive to potential inconsistencies in the  band-assignments resulting from crossings of bands with the same spin character.

Most importantly, our analyis of the NRSS and corresponding multipoles related to different structural distortion modes in LaVO$_3$, which is representative for the huge class of $Pbnm$ distorted perovskites, further revealed a more complex picture, suggesting that the total NRSS generally results from a superposition of different contributions of comparable magnitude related to multipoles of different rank. A quantitative theory of altermagnetism therefore needs to be based on a multi-dimensional order parameter containing multipoles of different rank as well as different components of the same rank.
Within this framework, the total NRSS can then be described as a sum over different channels, each related to a multipole of the charge and/or magnetization density, and the strength of each contribution is governed by a corresponding materials-specific altermagnetic coefficient. In future work, it might be instructive to isolate the contribution of each individual multipole, thus allowing for a characterization of altermagnetic materials based on their multipolar fingerprint.

\section*{Acknowledgments}
This work was supported by ETH Z\"{u}rich. Calculations were performed on the ETH Z\"{u}rich Euler cluster and the Swiss National Supercomputing Center Eiger cluster under Project ID s1304.

\bibliography{Altermagnetism}

\end{document}